\documentclass[12pt,draftclsnofoot,onecolumn]{IEEEtran}
\usepackage{amssymb}
\usepackage{amsmath}
\usepackage{cite}
\usepackage{algorithm}
\usepackage{algorithmic}
\usepackage[dvips]{graphicx}
\begin{document}
%
\title{Efficient Relay Beamforming Design with SIC Detection for Dual-Hop MIMO Relay Networks}

\author{Yu Zhang, Hanwen Luo, and Wen Chen, \emph{Member, IEEE}
\thanks{This work is supported by NSF China \#60972031, by SEU SKL project
\#W200907, by ISN project \#ISN11-01, by Huawei Funding
\#YJCB2009024WL and \#YJCB2008048WL,
and by National 973 project \#2009CB824900.}
\thanks{The authors are with the Electronic Engineering Department in Shanghai
        Jiao Tong University, Shanghai, China.
        Yu Zhang is also with
the State Key Laboratory of Integrated Services Networks, and Wen
Chen is also with SEU SKL for mobile communications (e-mail:
\{yuzhang49; hwluo; wenchen\}@sjtu.edu.cn).}}

\markboth{}{} \maketitle

\begin{abstract}
In this paper, we consider a dual-hop Multiple Input Multiple Output
(MIMO) relay wireless network, in which a source-destination pair
both equipped with multiple antennas communicates through a large
number of half-duplex amplify-and-forward (AF) relay terminals. Two
novel linear beamforming schemes based on the matched filter (MF)
and regularized zero-forcing (RZF) precoding techniques are proposed
for the MIMO relay system. We focus on the linear process at the
relay nodes and design the new relay beamformers by utilizing the
channel state information (CSI) of both backward channel and forward
channel. The proposed beamforming designs are based on the QR
decomposition (QRD) filter at the destination node which performs
successive interference cancellation (SIC) to achieve the maximum
spatial multiplexing gain. Simulation results demonstrate that the
proposed beamformers that fulfil both the intranode array gain and
distributed array gain outperform other relaying schemes under
different system parameters in terms of the ergodic capacity.

\end{abstract}

\begin{keywords}
MIMO relay, beamforming, successive interference cancellation, ergodic capacity.
\end{keywords}

\section{Introduction}
Recently relay wireless networks have drawn considerable interest
from both the academic and industrial communities. Due to
low-complexity and low-cost of the relay elements, the architectures
of multiple fixed relay nodes implemented in cellular systems and
many other kinds of networks are considered to be a promising
technique for future wireless networks~\cite{1}. Meanwhile, MIMO
technique is well verified to provide significant improvement in the
spectral efficiency and link reliability because of the multiplexing
and diversity gain~\cite{2,3}. Combining the relaying and MIMO
techniques can make use of  both advantages to increase the data
rate in the cellular edge and extend the network coverage.

The capacity of MIMO relay networks has been well investigated in
several papers~\cite{4,5,6}, in which, \cite{5} derives lower bounds
on the capacity of a Gaussian MIMO relay channel under the condition
of transmitting precoding. In order to improve the capacity of relay
networks, various kinds of linear distributed MIMO relaying schemes
have been investigated in~\cite{7,8,9,10,11,12,13,14}. In~\cite{7},
the authors analyze the stream signal-to-interference ratio
statistic and consider different relay beamforming based on the
finite-rate feedback of the channel states. Assuming
Tomlinson-Harashima precoding at the base station and linear
processing at the relay, \cite{8} proposes upper and lower bounds on
the achievable sum rate for the multiuser MIMO system with single
relay node. In~\cite{9}, a linear relaying scheme fulfilling the
target SNRs on different substreams is proposed and the
power-efficient relaying strategy is derived in closed form.
The optimal relay beamforming scheme and power control algorithms
for  a cooperative and cognitive radio system are presented
in~\cite{12}. In~\cite{13,14}, the authors design three relay
beamforming schemes based on matrix triangularization which have
superiority over the conventional zero-forcing (ZF) and
amplify-and-forward (AF) beamformers.

Inspired by these heuristic works, this paper proposes two novel
relay-beamformer designs for the dual-hop MIMO relay networks, which
can achieve both of the distributed array gain and intranode array
gain. Intranode array gain is the gain obtained from the
introduction of multiple antennas in each node of the dual-hop
networks. Distributed array gain results from the implementation of
multiple relay nodes and does not need any cooperation among them.
Assuming the same scenario given in~\cite{14}, the new relay
beamformers outperform the three schemes proposed in~\cite{14} under
various network conditions. The innovation points of our relaying
schemes are reflected in the matched filter and regularized
zero-forcing beamforming designs implemented at multiple relay nodes
while utilizing QRD of the effective channel matrix at the
destination node. The destination can perform SIC to decode multiple
data streams which have further enhancement effect on the channel
capacity.


In this paper, boldface lowercase letter and boldface uppercase
letter represent vectors and matrices, respectively. The notations
$\left( {\bf{A}} \right)_i$ and $\left( {\bf{A}} \right)_{i,j}$
represent the $i$th row and $(i,j)$th entry of the matrix
${\bf{A}}$. Notations $\textrm{tr}(\cdot)$ and $(\cdot)^H$ denote
trace and conjugate transpose operation of a matrix. Term
$\mbox{\boldmath $\mathbf{I}$}_N$ is an $N{\times}N$ identity
matrix. and $\|{\bf{a}}\|$ stands for the Euclidean norm of a vector
${\bf{a}}$. Finally, we denote the expectation operation by $\rm
E\{\cdot\}$.

\section{System Model}
The considered MIMO relay network consists of a single source and
destination node both equipped with $M$ antennas, and $K$
$N$-antenna relay nodes distributed between the source-destination
pair as illustrated in Fig. 1. When the source node implements
spatial multiplexing (SM), the requirement that $N \ge M$ must be
satisfied if every relay node is supposed to support all the $M$
independent data streams. We consider half-duplex non-regenerative
relaying throughout this paper where it takes two non-overlapping
time slots for the data to be transmitted from the source to the
destination node via the backward channel (BC) and forward channel
(FC). Due to deep large-scale fading effects produced by the long
distance, we assume that there is no direct link between the source
and destination. In this paper, the perfect CSIs of BC and FC are
assumed to be available at relay nodes. In a practical system, each
relay uses the training sequences or pilot sent from the source node
to acquire the CSI of all the backward channels. The acquisition
methods of FC's information would vary with two different duplex
forms. If it is a FDD system, the destination should estimate the
CSI of FC by using the relay-specific pilots first, and then
feedback the CSI to each relay node. As for a TDD system, due to its
intrinsic reciprocity, relay nodes can use the CSI of the link from
destination to relay nodes to acquire the CSI of FC.

In the first time slot, the source node broadcasts the signal to all
the relay nodes through BC. Let $M{\times}1$ vector $\mbox{\boldmath
$\mathbf{s}$}$ be the transmit signal vector satisfying the power
constraint ${\rm E}\left\{ {{\bf{ss}}^H } \right\} = \left( {{P
\mathord{\left/
 {\vphantom {P M}} \right.
 \kern-\nulldelimiterspace} M}} \right){\bf{I}}_M$, where $P$ is defined
as the total transmit power at the source node. Let ${\bf{H}}_k  \in
\mathbb{C}^{N \times M}$, $(k=1,...,K)$ stand for the BC MIMO
channel matrix from the source node to the $k$th relay node. All the
relay nodes are supposed to be located in a cluster. Then all the
backward channels ${\bf{H}}_1,\cdots,{\bf{H}}_K$ can be supposed to
be independently and identically distributed ($i.i.d.$) and
experience the same Rayleigh flat fading. Then the corresponding
received signal at the $k$th relay can be written as
\begin{equation}
{\bf{r}}_k  = {\bf{H}}_k {\bf{s}} + {\bf{n}}_k,
\end{equation}
where the term $\mbox{\boldmath$\mathbf{n}$}_{k}$ is the
spatio-temporally white zero-mean complex additive Gaussian noise
vector, independent across $k$, with the covariance matrix ${\rm
E}\left\{ {{\bf{n}}_k {\bf{n}}_k^H } \right\} = \sigma _1^2
{\bf{I}}_N$. Therefore, noise variance $\sigma _1^2$ represents the
noise power at each relay node.

In the second time slot, firstly each relay node performs linear
process by multiplying ${\bf{r}}_k$ with an $N \times N$ beamforming
matrix ${\bf{F}}_k$. Consequently, the signal vector sent from the
$k$th relay node is
\begin{equation}
{\bf{t}}_k  = {\bf{F}}_k {\bf{r}}_k.
\end{equation}
From more practical consideration, we assume that each relay node has its own power constraint satisfying
 ${\rm E}\left\{ {{\bf{t}}_k^H {\bf{t}}_k } \right\} \leq Q_k$,
 which is independent from power $P$. Hence
a power constraint condition of ${\bf{t}}_k$ can be derived as
\begin{equation}
p\left( {{\bf{t}}_k } \right) = tr\left\{ {{\bf{F}}_k \left(
{\frac{P}{M}{\bf{H}}_k {\bf{H}}_k^H  + \sigma _1^2 {\bf{I}}_N }
\right){\bf{F}}_k^H } \right\} \le Q_k.
\end{equation}
After linear relay beamforming process, all the relay nodes forward their data simultaneously to the destination.
Thus the signal vector received by the destination can be expressed as
\begin{align}
 {\bf{y}} & = \sum_{k=1}^{K}{\bf{G}}_k{\bf{t}}_k
+ {\bf{n}}_d  =  \sum_{k=1}^{K} {\bf{G}}_k {\bf{F}}_k  {\bf{H}}_k
{\bf{s}} +\sum_{k=1}^{K} {\bf{G}}_k {\bf{F}}_k {\bf{n}}_k
+{\bf{n}}_d,
\end{align}
where ${\bf{G}}_k$, under the same assumption as ${\bf{H}}_k$, is the $M \times N$ forward channel
between the $k$th relay node and the destination. ${\bf{n}}_d \in \mathbb{C}^M$, satisfying
${\rm E}\left\{ {{\bf{n}}_d {\bf{n}}_d^H } \right\} = \sigma _2^2 {\bf{I}}_M$,
denotes the zero-mean white circularly symmetric complex additive Gaussian noise at the destination
node with the noise power $\sigma _2^2$.

\section{Relay Beamforming Design}
In this section, the network ergodic capacity with the QR detector
applied at the destination node for SIC detection is analyzed.
And then we will propose two novel relay beamformer schemes based on the
MF and RZF beamforming techniques.

\subsection{QR decomposition and SIC detection}
Conventional receivers such as MF, zero-forcing (linear decorrelator) and
linear minimum mean square error (L-MMSE) decoder have been well studied in the previous works.
Matched filter receiver has bad performance in the high SNR region while ZF produces noise enhancement
effect. MMSE equalizer which can be seen as a good tradeoff of the MF and ZF receivers, however,
achieves the same order of diversity as ZF does. Hence much larger intranode array gain also cannot be
obtained from the MMSE receiver.
As analyzed in [15], SIC detection based on the QRD has significant advantage over
those conventional detectors and the performance of the QR detector is asymptotically equivalent to
that of the maximum-likelihood detector (MLD). So we will utilize the QRD detector
as the destination receiver $\bf{W}$ throughout this paper.

From the above discussion, the final received signal at destination can be derived as follows.
Let the term $\sum_{k=1}^{K}{\bf{G}}_k{\bf{F}}_k {\bf{H}}_k={\bf{H}}_{\mathcal {S}\mathcal {D}} $, and
$\sum_{k=1}^{K}{\bf{G}}_k {\bf{F}}_k{\bf{n}}_k
+{\bf{n}}_d = {\bf{z}}$. Then equation (4) can be rewritten as
\begin{equation}
 {\bf{y}}= {\bf{H}}_{\mathcal {S}\mathcal {D}} {\bf{s}}
+ {\bf{z}},
\end{equation}
where ${\bf{H}}_{\mathcal {S}\mathcal {D}}$ represents the effective channel between the source
and destination node, and ${\bf{z}}$ is the effective noise vector cumulated from
the noise $\bf{n}_k$ at each relay node
and the noise vector ${\bf{n}}_d$ at the destination.
Implement QR decomposition of the effective channel as
\begin{equation}
  {\bf{H}}_{\mathcal {S}\mathcal {D}}={\bf{Q}}_{\mathcal {S}\mathcal {D}} {\bf{R}}_{\mathcal {S}\mathcal {D}},
\end{equation}
where ${\bf{Q}}_{\mathcal {S}\mathcal {D}}$ is an $M \times M$ unitary matrix and ${\bf{R}}_{\mathcal {S}\mathcal {D}}$
is an $M \times M$ right upper triangular matrix. Therefore the QR detector at destination node is chosen as: ${\bf{W}}={\bf{Q}}_{\mathcal {S}\mathcal {D}}^H $,
and the signal vector after detection becomes
\begin{equation}
 \tilde {\bf{y}}= {\bf{R}}_{\mathcal {S}\mathcal {D}} {\bf{s}}
+ {\bf{Q}}_{\mathcal {S}\mathcal {D}}^H {\bf{z}}.
\end{equation}
Finally, the optimal relay beamformer design problem can be formulated mathematically as
\begin{align}
 & {\bf{\hat F}}_k  = \arg \mathop {\max }\limits_{{\bf{F}}_k } \;C\left( {{\bf{F}}_k } \right),\\
 & s.t. \quad {\rm{ }}p\left( {{\bf{t}}_k } \right) \le Q_k {\rm{ }},
\end{align}
 where $C\left( {{\bf{F}}_k } \right)$ is the network ergodic capacity having various specific forms
 decided by destination detector ${\bf{W}}$ and relay beamforming matrix ${\bf{F}}_k$ that will be discussed in detail in the following subsections.

Note that the closed-form solution is difficult to obtain when
trying to solve the optimization problem (8) directly. In order to
get a specific form of the relay beamformers, we further assume that
a power control factor $\rho_k $ is set with $\bf{F}_k$ in (2) to
guarantee that each relay transmit power is equal to $Q_k$. Since
$\textbf{H}_1,\cdots,\textbf{H}_K$ (and
$\textbf{G}_1,\cdots,\textbf{G}_K$) are $i.i.d.$ distributed and
experience the same Rayleigh fading, all the relay beamformers can
have a uniform design type. Hence the transmit signal from each
relay node after linear beamforming and power control becomes
\begin{equation}
 {\bf{t}}_k  = \rho_k {\bf{F}}_k {\bf{r}}_k,
\end{equation}
where the power control parameter $\rho_k$ can be derived from equation (3) as
\begin{equation}
 \rho _k  = \left( {{Q_k \mathord{\left/
 {\vphantom {Q {tr\left\{ {{\bf{F}}_k^H \left( {\frac{P}{M}{\bf{H}}_k {\bf{H}}_k^H  + \sigma _1^2 {\bf{I}}_N } \right){\bf{F}}_k } \right\}}}} \right.
 \kern-\nulldelimiterspace} {tr\left\{ {{\bf{F}}_k \left( {\frac{P}{M}{\bf{H}}_k {\bf{H}}_k^H  + \sigma _1^2 {\bf{I}}_N } \right){\bf{F}}_k^H } \right\}}}} \right)^{\frac{1}{2}}.
\end{equation}

\subsection{MF beamforming}
According to the principles of maximum-ratio-transmission (MRT) [16] and maximum-ratio-combining (MRC) [17],
we choose the MF as the beamformer for each relay node. Therefore we get the beamforming matrix as
\begin{equation}
{\bf{F}}^{MF}_k  =  {\bf{G}}_k^H {\bf{H}}_k^H,
\end{equation}
where each relay beamformer can be divided into two parts: a receive
beamformer ${\bf{H}}_k^H$ and a transmit beamformer ${\bf{G}}_k^H$.
The receive beamformer ${\bf{H}}_k^H$ is the optimal weight matrix
that maximizes received SNR at the relay. Consequently, the received
signal at the destination can be rewritten from (10) and (12) as
\begin{equation}
{\bf{y}} = \underbrace {\sum\limits_{k = 1}^K {\rho _k {\bf{G}}_k } {\bf{G}}_k^H {\bf{H}}_k^H {\bf{H}}_k }_{{\bf{H}}_{{\cal S}{\cal D}}^{MF} }{\bf{s}} + \underbrace {\sum\limits_{k = 1}^K {\rho _k {\bf{G}}_k } {\bf{G}}_k^H {\bf{H}}_k^H {\bf{n}}_k  + {\bf{n}}_d }_{{\bf{z}}^{MF} },
\end{equation}
where $\rho_k$ is given by substituting (12) into equation (11).
Performing QRD of the  ${\bf{H}}^{MF}_{\mathcal {S}\mathcal {D}}$ as
\begin{equation}
  {\bf{H}}^{MF}_{\mathcal {S}\mathcal {D}}={\bf{Q}}^{MF}_{\mathcal {S}\mathcal {D}} {\bf{R}}^{MF}_{\mathcal {S}\mathcal {D}}.
\end{equation}
Then we get the destination receiver as
\begin{equation}
{\bf{W}}^{MF}= \left( {{\bf{Q}}_{\mathcal {S}\mathcal {D}}^{MF} } \right)^H.
\end{equation}
Hence the signal vector after QR detection becomes
\begin{equation}
 \tilde {\bf{y}}^{MF}= {\bf{R}}^{MF}_{\mathcal {S}\mathcal {D}} {\bf{s}}
+ \left( {{\bf{Q}}_{\mathcal {S}\mathcal {D}}^{MF} } \right)^H {\bf{z}}^{MF}.
\end{equation}
Note that the matrix ${\bf{R}}^{MF}_{\mathcal {S}\mathcal {D}}$ has the right upper triangular form as
\begin{equation}
  {\bf{R}}^{MF}_{\mathcal {S}\mathcal {D}}  = \left( {\begin{array}{*{20}c}
   {r_{1,1} } & {r_{1,2} } &  \ldots  & {r_{1,M} }  \\
   {} & {r_{2,2} } & {} &  \vdots   \\
   {} & {} &  \ddots  & {}  \\
   {\bf{0}} & {} & {} & {r_{M,M} }  \\
\end{array}} \right),
\end{equation}
where the diagonal entries $r_{m,m} \,(m=1,...,M)$ of (17) are real positive numbers.
With the destination node carrying out the SIC detection, the effective
SNR for the $m$th data stream of MF relay beamforming scheme can be derived as
\begin{equation}
SNR_m^{MF}  = \frac{{\left( {{P \mathord{\left/
 {\vphantom {P M}} \right.
 \kern-\nulldelimiterspace} M}} \right)r_{m,m}^2 }}{{\left( {\sum\limits_{k = 1}^K {\left\| {\left( {\rho _k \left( {{\bf{Q}}_{\mathcal {S}\mathcal {D}}^{MF} } \right)^H {\bf{G}}_k {\bf{G}}_k^H {\bf{H}}_k^H } \right)_m } \right\|^2 } } \right)\sigma _1^2  + \sigma _2^2 }}.
\end{equation}

%

\subsection{MF-RZF beamforming}
In this subsection, we utilize the regularized zero-forcing (RZF)
precoding [18] as the transmit beamformer for FC while MF is still
kept as the receive beamformer matching with the BC condition. So
the MF-RZF beamformer is constructed as
\begin{equation}
{\bf{F}}_k^{MF - RZF}  =  {\bf{G}}_k^H \left( {{\bf{G}}_k {\bf{G}}_k^H  + \alpha_k {\bf{I}}_M } \right)^{ - 1} {\bf{H}}_k^H,
\end{equation}
where $\alpha_k$ is an adjustable parameter that controls the amount of interference
among multiple data streams in the second hop. One possible metric for choosing
$\alpha_k$ is to maximize the end-to-end effective SNR which will be given below.
Hence the corresponding received signal at the destination is
\begin{align}
  {\bf{y}} & = \sum_{k=1}^{K} {\bf{G}}_k {\bf{F}}_k  {\bf{H}}_k {\bf{s}}
+\sum_{k=1}^{K} {\bf{G}}_k {\bf{F}}_k {\bf{n}}_k \nonumber
+{\bf{n}}_d \nonumber \\
& = \sum\limits_{k = 1}^K {\rho _k {\bf{G}}_k } {\bf{G}}_k^H \left(
{{\bf{G}}_k {\bf{G}}_k^H  + \alpha_k {\bf{I}}_M } \right)^{ - 1}
{\bf{H}}_k^H {\bf{H}}_k {\bf{s}} + \sum\limits_{k = 1}^K {\rho _k
{\bf{G}}_k } {\bf{G}}_k^H \left( {{\bf{G}}_k {\bf{G}}_k^H + \alpha_k
{\bf{I}}_M } \right)^{ - 1} {\bf{H}}_k^H {\bf{n}}_k  + {\bf{n}}_d.
\end{align}
The effective channel matrix between the source and the destination
is derived from (20) as
\begin{equation}
{\bf{H}}^{MF-RZF}_{\mathcal {S}\mathcal {D}}  = \sum\limits_{k = 1}^K {\rho _k {\bf{G}}_k } {\bf{G}}_k^H \left( {{\bf{G}}_k {\bf{G}}_k^H  + \alpha_k {\bf{I}}_M } \right)^{ - 1} {\bf{H}}_k^H {\bf{H}}_k.
\end{equation}
Similarly, after QRD of ${\bf{H}}^{MF-RZF}_{\mathcal {S}\mathcal {D}}$ and
the SIC detection at the destination node, the effective SNR for the $m$th
data stream of MF-RZF relay beamforming is obtained as
\begin{equation}
SNR_m^{MF-RZF}  = \frac{{\left( {{P \mathord{\left/
 {\vphantom {P M}} \right.
 \kern-\nulldelimiterspace} M}} \right)\tilde r_{m,m}^2 }}{{\left( {\sum\limits_{k = 1}^K {\left\| {\left( {\rho _k \left( {{\bf{Q}}_{SD}^{MF - RZF} } \right)^H {\bf{A}}_k } \right)_m } \right\|^2 } } \right)\sigma _1^2  + \sigma _2^2 }},
\end{equation}
where $ {\bf{A}}_k = {\bf{G}}_k {\bf{F}}_k^{MF-RZF}$. Term $\tilde
r_{m,m}$ is the $m$th diagonal entry of the right upper triangular
matrix ${\bf{R}}^{MF-RZF}_{\mathcal {S}\mathcal {D}}$ derived from
QRD operation of ${\bf{H}}^{MF-RZF}_{\mathcal {S}\mathcal {D}}$ like
(14). And $\rho_k$ of the MF-RZF relay beamforming is given by
substituting (19) into equation (11).

%

Finally, the ergodic capacity of a dual-hop MIMO relay network with relay beamforming can be derived by summing up
the data rate of all the streams as
\begin{equation}
C = E_{\left\{ {{\bf{H}}_k ,{\bf{G}}_k } \right\}_{k = 1}^K } \left\{ {\frac{1}{2}\sum\limits_{m = 1}^M {\log _2 \left( {1 + SNR_m } \right)} } \right\},
\end{equation}
where $SNR_m $ refers to the effective SNR in (18) or (22). From the
cut-set theorem in network information theory [6], the upper bound
capacity of the MIMO relay networks is
\begin{equation}
C_{upper}  = \rm E_{\left\{ {{\bf{H}}_k } \right\}_{k = 1}^K } \left\{ {\frac{1}{2}\log \det \left( {{\bf{I}}_M  + \frac{P}{{M \sigma _1^2 }}\sum\limits_{k = 1}^K {{\bf{H}}_k^H {\bf{H}}_k } } \right)} \right\}.
\end{equation}

\subsection{Computational complexity analysis and remarks}

In spite of no additional signal processing at the destination,
referenced schemes in [14] implement QR decomposition of matrices at
each relay node actually. More precisely, for QR-P-QR scheme in
[14], each backward channel ${\bf{H}}_k$ and forward channel
${\bf{G}}_k^H$ should have a QRD operation. Each relay node has
twice QRD operations of $N \times M$ complex matrix. Therefore, it
costs $2K$ times of QRD ($N \times M$ complex matrix) for QR-P-QR
scheme. For QR-P-ZF scheme, it still needs to implement $K$ times of
QRD of the $N \times M$ matrix. When it comes to our schemes, for
both MF and MF-RZF relay beamforming, the whole signal processing
spend only once QRD at the destination node. Moreover, in our design
the QRD is operated on the effective channel matrix
${\bf{H}}_{\mathcal {S}\mathcal {D}}$ between the source and the
destination. The dimension of the complex matrix for QRD is $M
\times M$, which is free from the antenna number $N$ and the relay
number $K$. Obviously, the proposed schemes reduce the computational
complexity sharply compared with the referenced methods in [14].

Additionally, in order to guarantee the effective channel matrix to
take the right lower triangular form, the phase control and ordering
matrix has to be used in the relay beamformers in [14]. This results
in a performance loss in terms of the network capacity. While the
QRD of the compound effective channel at the destination proposed in
this paper makes the relay beamformer design more flexible, because
the effective channel matrix is not necessary to be a triangular
form.

\section{Simulation Results}
In this section, numerical simulations are carried out in order to
verify the performance superiority of the proposed relay beamforming
strategies. We compare the ergodic capacities of MF and MF-RZF relay
beamformers with QR-P-QR, QR-P-ZF proposed in [14] and the
conventional AF relaying scheme in the dual-hop MIMO relay networks.
The capacity upper bound is also taken into account as a baseline.
All the schemes are compared under the condition of various system
parameters including total number of relay nodes and power
constraints at source and relay nodes, i.e., different PNR
($P/\sigma_1^2$, the SNR of BC), and different QNR
($Q_k/\sigma_2^2$, the SNR of FC). For simplicity, the exntries of
${{\bf{H}}_k }$ and ${{\bf{G}}_k }$ are assumed to be $i.i.d.$
complex Gaussian with zero mean and unit variance. All the relay
nodes are supposed to have the same power constraint $Q_k=Q
\,(k=1,...,K)$, and $\alpha_k=1 \,(k=1,...,K)$, which, within a
limited range, has no significant impact on the ergodic capacity of
the MF-RZF relay beamforming.

\subsection{Capacity versus Total Number of Relay Nodes}
Like in~\cite{13,14}, the capacity comparisons are given with the
increase of the total number of relay nodes. In order to illustrate
how the SNRs of BC and FC have impact on  the ergodic capacity with
various relay beamforming schemes, three different PNR and QNR are
taken into account. Fig.~2 shows the capacities change with $K$ when
$N=M=4, \mbox{PNR}=\mbox{QNR}=10$dB. Apparently, the proposed MF and
MF-RZF relay beamformers outperform the QR-P-ZF and QR-P-QR relaying
schemes~in [14] for $K>1$. For this moderate PNR and QNR, the MF-RZF
beamformer has the best ergodic capacity performance among the five
relaying schemes and approaches to the capacity upper bound. This
can be explained as a result that the MF receive beamformer can
maximize receive SNRs at each relay node while the RZF transmit
beamformer pre-cancel inter-stream interference before transmitting
the signal to the destination node.

The relative capacity gains changing with the PNR and QNR is
demonstrated in Fig. 3 and Fig. 4. It can be seen from Fig.~3 that
MF and MF-RZF keep the superiority over other relaying schemes when
the network has low SNR in BC (PNR$=5$dB) and high SNR in FC
(QNR$=20$dB). This is because that the MF is used as the receive
beamformer for the first hop channel, showing the advantage of MF
against the low SNR condition. Furthermore, Fig.~3 shows that the
capacity gains of MF-RZF scheme over other beamformers become
larger, while the performance superiority of MF decreases when
compared to the scenario in Fig.~2. This is because that the MF
performance becomes worse with the increase of SNR, while the RZF in
FC turns to be better. A larger gap between QR-P-ZF and QR-P-QR
beamforming schemes also confirms the advantage of ZF being the
transmit beamformer in the high SNR region. With the knowledge of
the performance characteristics of MF in low SNR regions and RZF in
high SNR regions, the fact illustrated in Fig. 4 that ergodic
capacity of MF-RZF becomes a little bit smaller than MF in low QNR
environment is reasonable.

Finally, in all the three environments considered above, the
conventional AF relaying keeps as a bad relaying strategy. It can be
seen that AF can not obtain the distributed array gain since its
ergodic capacity does not increase with the total number of relay
nodes. The reason is that, as for the AF relaying, each relay node
uses the identity matrix as the beamformer which does not utilize
any CSI of both BC and FC. It is also very important to investigate
the behaviors of all the relay beamforming schemes when distributed
array gain is unavailable, i.e., when there is only a single relay
node in the network. From Fig.~2 to Fig.~4, it can be seen that AF
relaying is no longer the worst one and becomes acceptable when
$K=1$. Meanwhile, the performance advantages of the proposed methods
over other conventional schemes vary from case to case. Look at the
ergodic capacities of all the schemes at the point of $K=1$ in
Fig.~3. At this time, the single relay system has low PNR
(PNR$=5$dB)and high QNR (QNR$=10$dB). MF-RZF's capacity has about
$0.1$bps loss than QR-P-ZF beamforming while MF has $0.03$bps gain
over QR-P-QR scheme. However, if the dual-hop network has moderate
PNR and QNR (see Fig.~2) or high QNR (see Fig.~4), the MF and MF-RZF
still outperform the schemes proposed in [14]. For example, when
$K=1, \mbox{PNR}=\mbox{QNR}=10$dB, the ergodic capacity of MF-RZF
beamforming achieves $0.3$bps and $1.01$bps gains over QR-P-QR and
QR-P-ZF schemes respectively. As for the MF beamformer, these gains
become $0.05$bps and $0.77$bps. From the above discussion, it can be
concluded that our proposed relaying schemes are still efficient
when the relay network has no distributed array condition and only
intranode array gain is available. It should be noticed that
simplest AF relaying has desirable capacity performance in this
case. Therefore, the AF scheme might be regarded as an alternative
solution, especially when the network has only one relay node and
moderate SNRs of two-hop channels.

\subsection{Capacity versus PNR}
The ergodic capacity versus the PNR and QNR is another important
aspect to measure the performance of the proposed schemes. The
performances of MF and MF-RZF linear relaying schemes are shown in
Fig.~5 and Fig.~6. We set QNR$=$PNR in Fig.~5, which is the same as
done in~[14]. The ergodic capacities of both MF-RZF and MF relaying
strategies grow approximately linearly with the PNR (and QNR) like
the upper bound and outperform other schemes.

In Fig.~6, we evaluate how the capacities change with the PNR by
keeping QNR$=10$dB. The two proposed relay beamformers can still
achieve much better performance than the conventional schemes.
However, the ergodic capacities of all the relay beamforming schemes
become saturated as the PNR increases. Note that AF scheme can even
outperform the QR-P-ZF beamforming in the high PNR region in this
case. And capacity upper bound keeps growing linearly with PNR since
it is determined only by the BC conditions as can be verified in
equation (24). The result in Fig.~6 illustrates that if the SNR of
FC keeps under certain values, simply increasing the source transmit
power has limited impact on the network capacity.

\section{Conclusion and Future Work}
In this paper, two novel relay beamformer design schemes based on MF
and RZF techniques have been derived for a dual-hop MIMO relay
network with Amplify-and-Forward (AF) relaying protocol. The
proposed MF and MF-RZF beamformers are constructed jointly with the
QR decomposition filer at the destination node which transforms the
effective compound channel into a right upper triangular form.
Consequently multiple data streams can be decoded with the
destination SIC detector. Simulation results demonstrate that our
proposed schemes outperform the conventional relay beamforming
strategies in the sense of the ergodic capacity under various
network parameters. Furthermore, the two proposed relay beamforming
schemes still have desirable performance when the distributed array
gain is unavailable in the network.

Although the proposed relay beamforming strategies have performance
gain over the conventional schemes, the original optimization
problem (8) and (9), the imperfect CSIs of BC and FC, the overhead
of the feedback traffic, and the optimal $\alpha_k$ values of the
MF-RZF beamformer are still challenging problems that need further
research effort.

%


\begin{figure}[htp]
\centering
\includegraphics[width=6in]{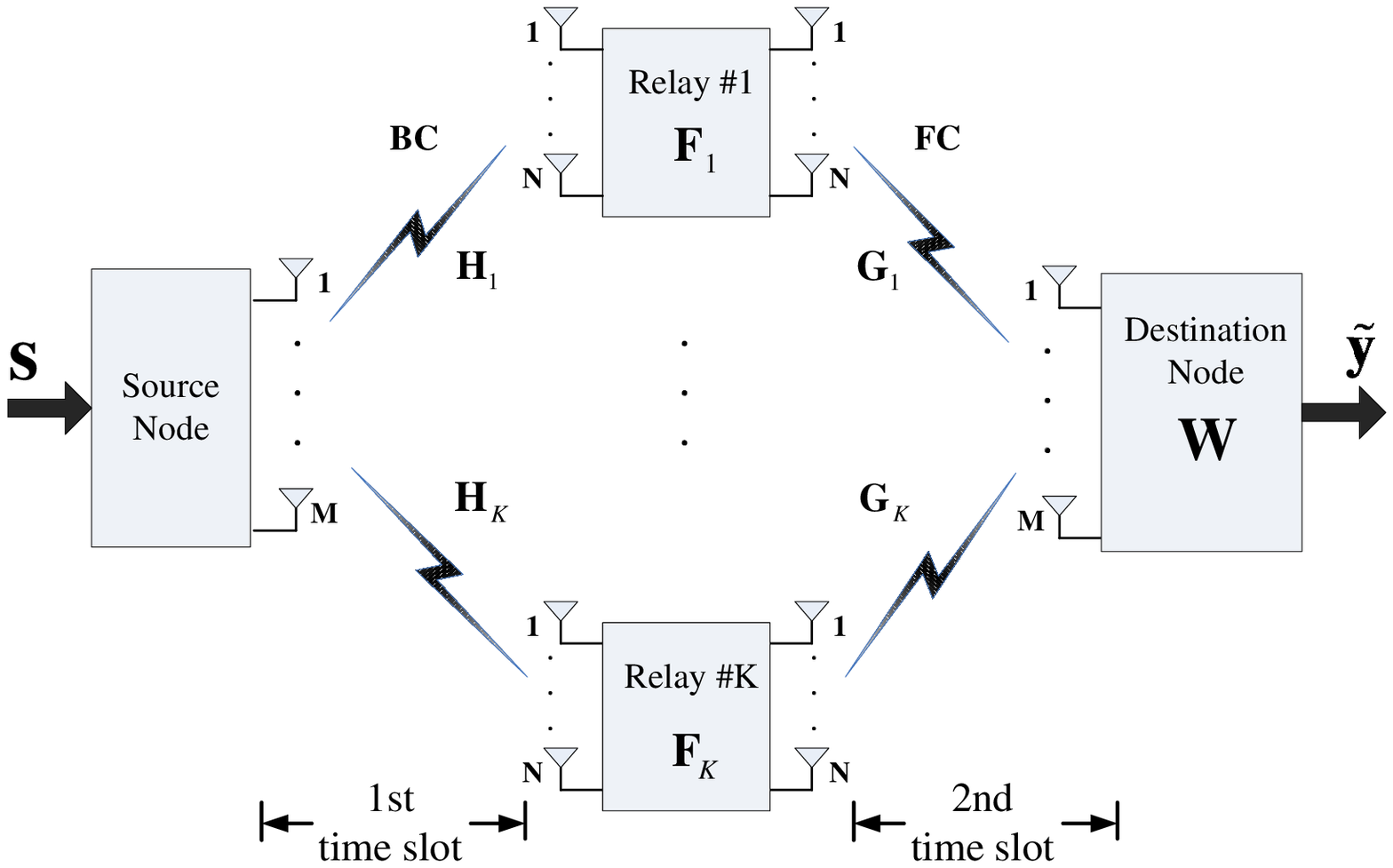}
\caption{System model of a dual-hop MIMO network with relay beamforming.}
\end{figure}

\begin{figure}
\centering
\includegraphics[width=6in]{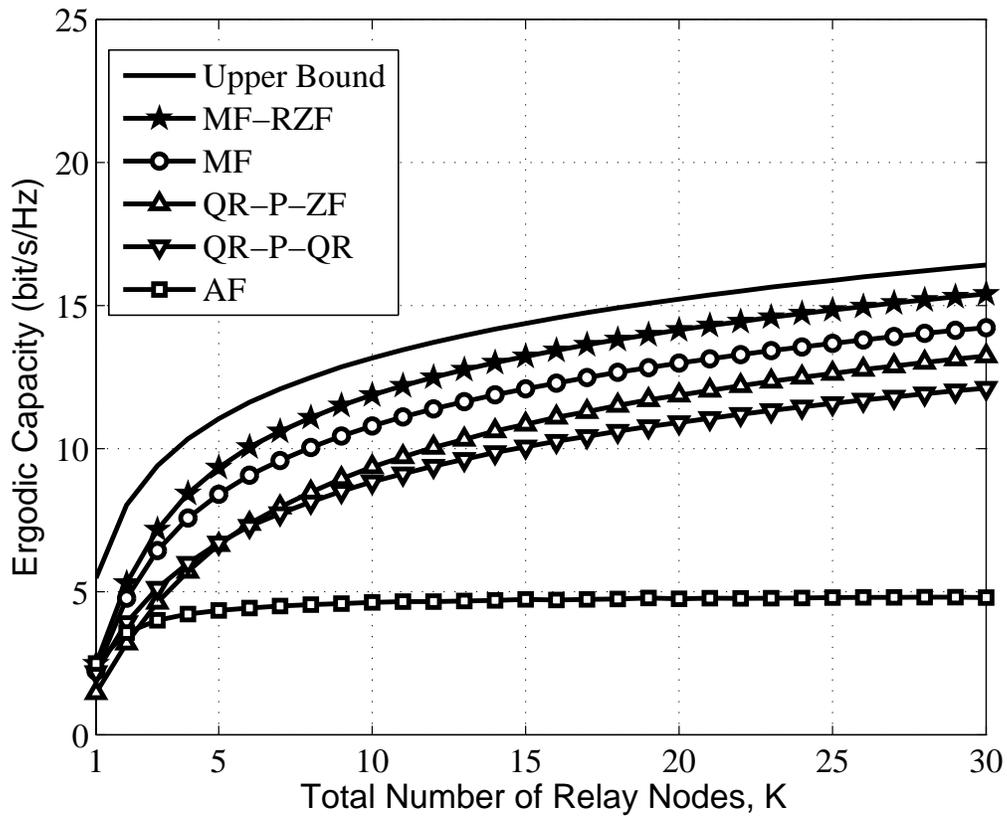}
\caption{Ergodic capacity comparisons versus $K$ ($N=M=4, PNR=QNR=10dB$).}
\end{figure}

\begin{figure}
\centering
\includegraphics[width=6in]{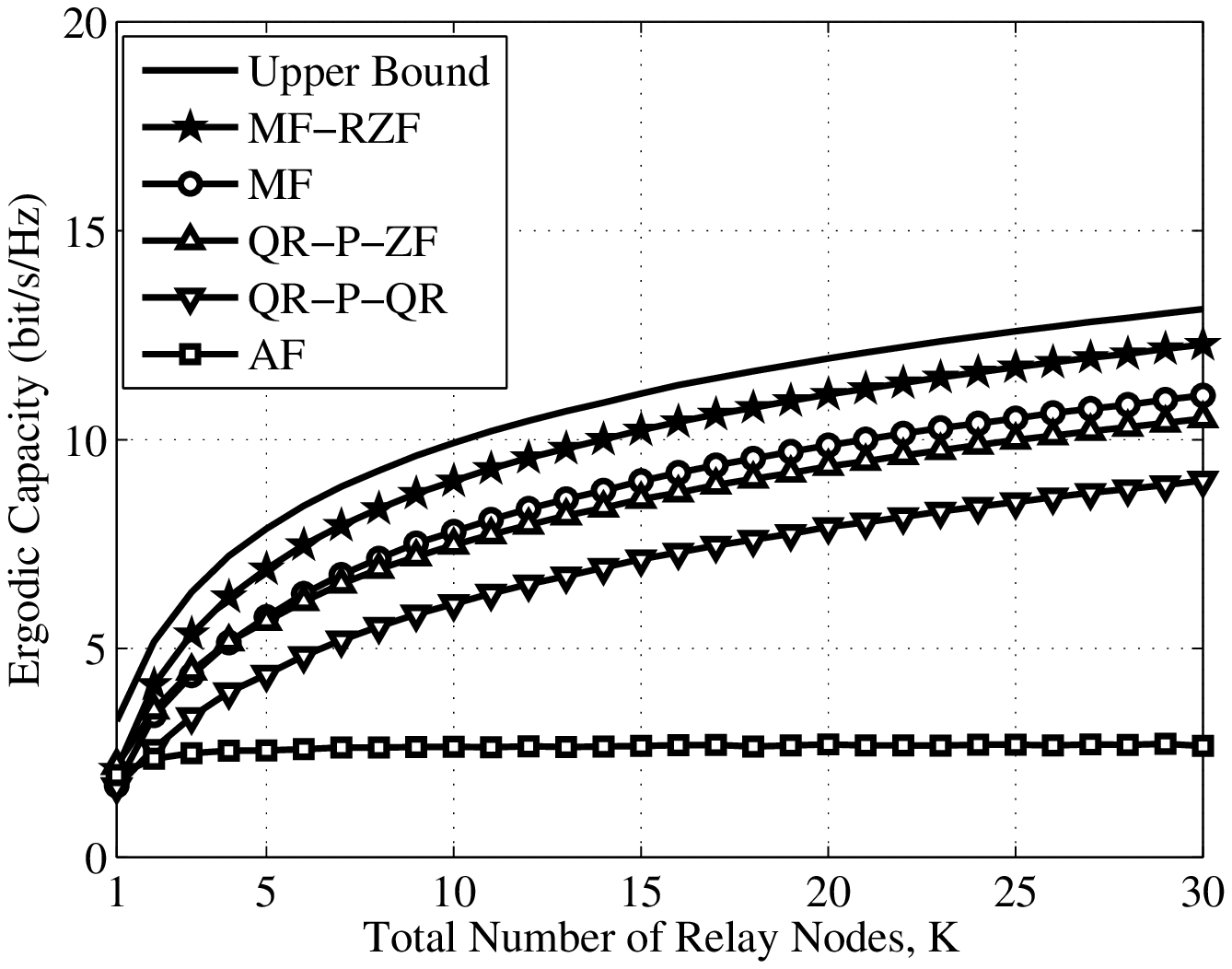}
\caption{Ergodic capacity comparisons versus $K$ ($N=M=4, PNR=5dB, QNR=20dB$).}
\end{figure}

\begin{figure}
\centering
\includegraphics[width=6in]{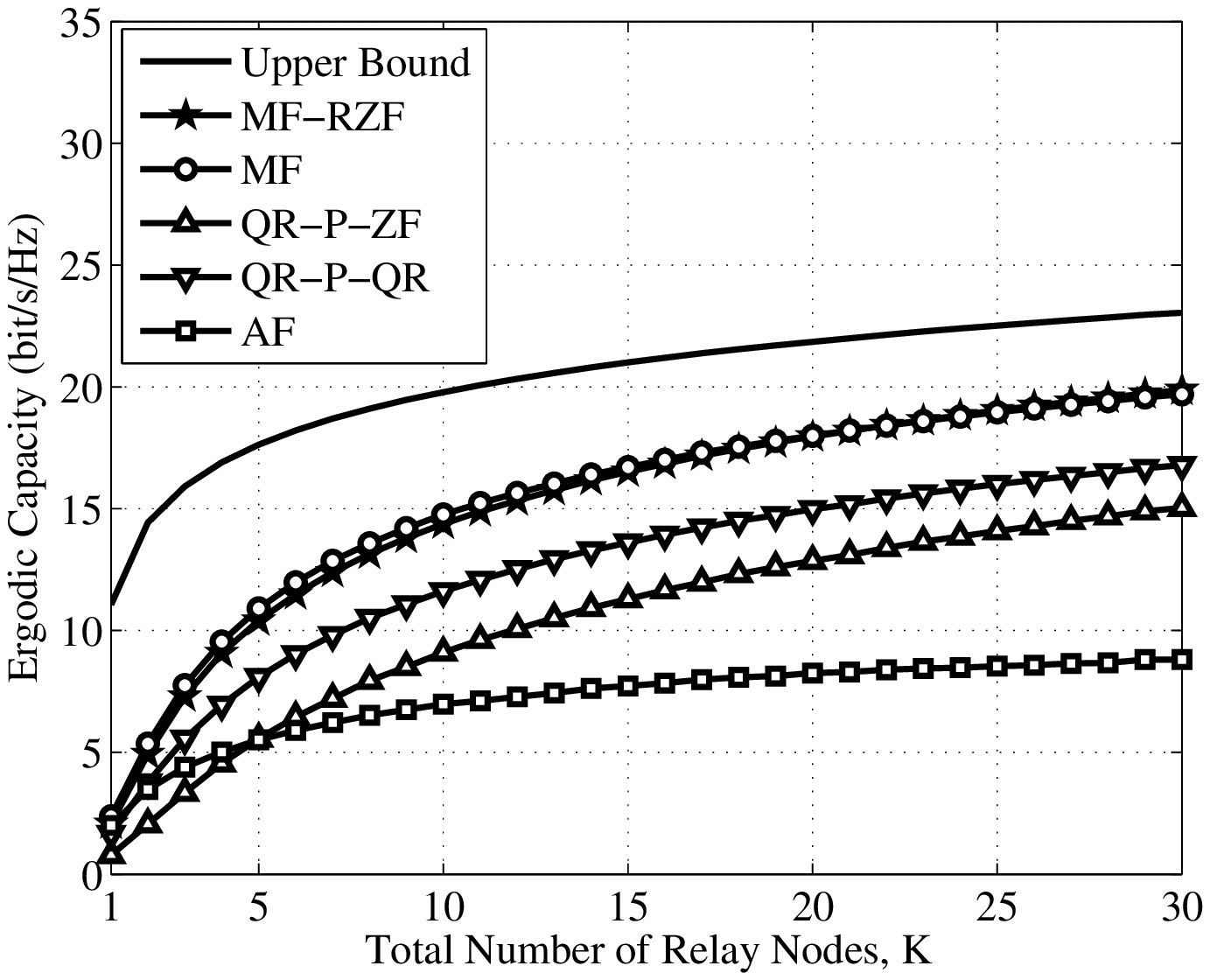}
\caption{Ergodic capacity comparisons versus $K$ ($N=M=4, PNR=20dB, QNR=5dB$).}
\end{figure}

\begin{figure}
\centering
\includegraphics[width=6in]{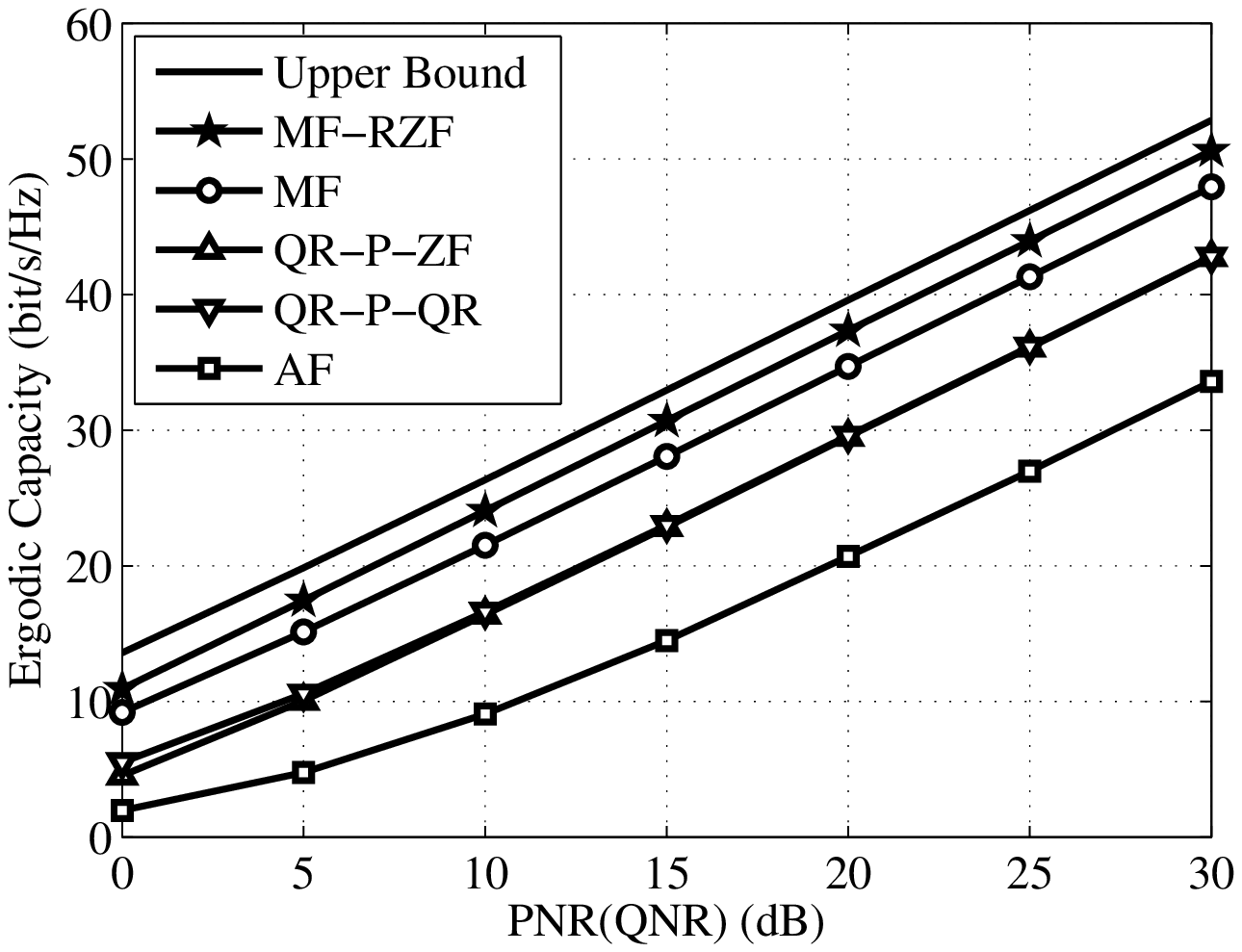}
\caption{Ergodic capacity comparisons versus PNR (QNR) ($N=M=8, K=10$).}
\end{figure}

\begin{figure}
\centering
\includegraphics[width=6in]{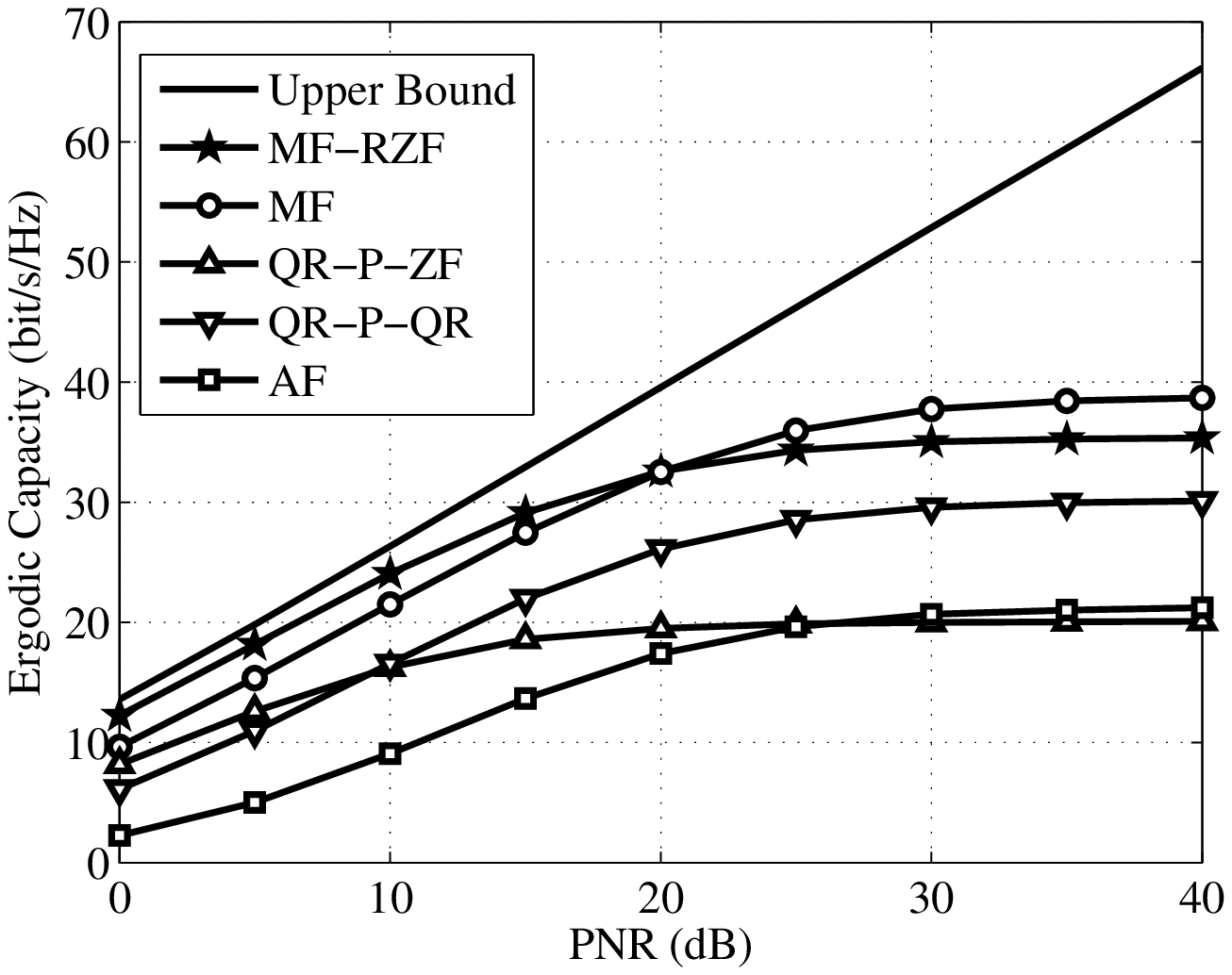}
\caption{Ergodic capacity comparisons versus PNR ($N=M=8, QNR=10dB, K=10$).}
\end{figure}


\end{document}